\documentclass[sigconf,screen]{acmart}

\usepackage{enumitem}
\usepackage[linesnumbered,ruled,vlined]{algorithm2e}
\usepackage{xcolor}
\usepackage{subfigure}
\usepackage{multirow}
\usepackage{makecell}
\usepackage{stmaryrd}

\newcommand{\etal}{{\textit{et al.}}\xspace}
\newcommand{\ie}{{\textit{i.e.}}\xspace}
\newcommand{\eg}{{\textit{e.g.}}\xspace}

\newcommand{\wasm}{Wasm\xspace}
\newcommand{\name}{\texttt{WarpL}\xspace}
\newcommand{\codes}{\textit{slow code}\xspace}

\AtBeginDocument{%
  }

\copyrightyear{2026}
\acmYear{2026}
\setcopyright{cc}
\setcctype{by}
\acmConference[ICSE '26]{2026 IEEE/ACM 48th International Conference on Software Engineering}{April 12--18, 2026}{Rio de Janeiro, Brazil}
\acmBooktitle{2026 IEEE/ACM 48th International Conference on Software Engineering (ICSE '26), April 12--18, 2026, Rio de Janeiro, Brazil}
\acmPrice{}
\acmDOI{10.1145/3744916.3773141}
\acmISBN{979-8-4007-2025-3/26/04}

\begin{document}

\title{Debugging Performance Issues in WebAssembly Runtimes via Mutation-based Inference}

\author{Ruiying Zeng}
\orcid{0009-0000-6367-0077}
\affiliation{
  \department{College of Computer Science \\ and Artificial Intelligence}
  \institution{Fudan University \\
 \& Shanghai Key Laboratory of \\ Intelligent Information Processing}
  \city{Shanghai}
  \country{China}
}
\email{ryzeng22@m.fudan.edu.cn}

\author{Shuyao Jiang}
\orcid{0000-0002-2797-875X}
\affiliation{
  \department{Department of Computer Science \\ and Engineering}
  \institution{The Chinese University of Hong Kong}
  \city{Hong Kong}
  \country{China}
}
\email{syjiang21@cse.cuhk.edu.hk}

\author{Wenxuan Zhao}
\orcid{0009-0005-9574-3310}
\affiliation{%
  \department{College of Computer Science \\ and Artificial Intelligence}
  \institution{Fudan University \\
 \& Shanghai Key Laboratory of \\ Intelligent Information Processing}
  \city{Shanghai}
  \country{China}
}
\email{wxzhao22@m.fudan.edu.cn}

\author{Yangfan Zhou}
\authornote{Yangfan Zhou is the corresponding author.}
\orcid{0000-0002-9184-7383}
\affiliation{%
  \department{College of Computer Science \\ and Artificial Intelligence}
  \institution{Fudan University \\
 \& Shanghai Key Laboratory of \\ Intelligent Information Processing}
  \city{Shanghai}
  \country{China}
}
\email{zyf@fudan.edu.cn}


\begin{abstract}
Performance debugging in WebAssembly (\wasm) runtimes is essential for ensuring the robustness of \wasm, especially since performance issues have frequently occurred in \wasm runtimes, which can significantly degrade the capabilities of hosted services.
Many performance issues in \wasm runtimes result from suboptimal compilation of input \wasm programs, for which existing performance debugging methods primarily designed for application-level inefficiencies are not well-suited.
In this paper, we present \name, a novel mutation-based approach that aims to identify the exact suboptimal instruction sequences responsible for the performance issues in \wasm runtimes, thereby narrowing down the root causes.
Specifically, \name obtains a functionally similar mutant in which the performance issue does not manifest, and isolates the exact suboptimal instructions by comparing the machine code of the original and mutated programs.
We implement \name as an open-source tool and evaluate it on 12 real-world performance issues across three widely used \wasm runtimes.
\name identified the exact causes in 10 out of 12 issues.
Notably, we have used \name to successfully diagnose six previously unknown performance issues in Wasmtime.
\end{abstract}



\received{18 July 2025}
\received[accepted]{17 October 2025}

\maketitle

\section{Introduction}\label{sec_intro}

Performance is of critical concern for WebAssembly (abbreviated \wasm) ~\cite{haas2017bringing} runtimes toward mature \wasm applications, especially since \wasm has been enthusiastically advocated as a good solution for hosting cloud services due to its portability and memory safety~\cite{simon2020Faasm,philipp2022pushing,Phani2020Sledge}.
Recent studies have found that performance issues (\ie, abnormal latency) frequently occur in \wasm runtimes~\cite{abhinav2019not,jiang2023revealing,jiang2025distinguishability}.
Such issues can lead to severe degradation of the service capability.
A previous study~\cite{jiang2023revealing} revealed that a 30ms-latency could lead to a throughput reduction of up to 50\% in real-world \wasm microservices~\cite{microservice}.
Thus, \textit{performance debugging} (\ie, root cause diagnosis of performance issues) in \wasm runtimes is crucial to the robustness of \wasm.

\begin{figure}
    \centering
    \includegraphics[width=1\linewidth]{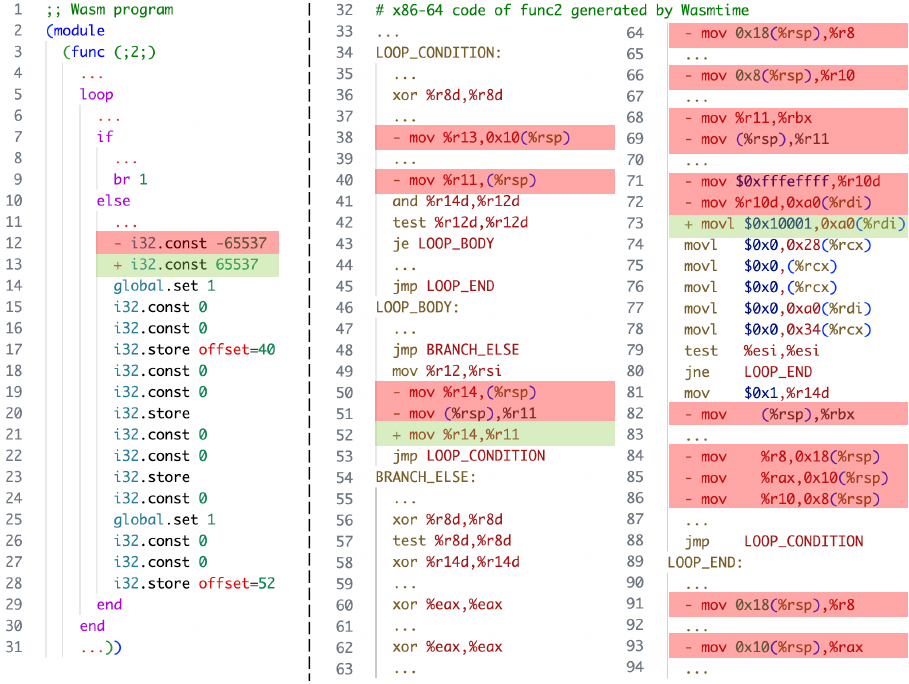}
    \caption{\textbf{A case causing a long execution time in Wasmtime. The differences between the original \wasm bytecode (red) and its mutant (green) are shown in both \wasm bytecode and the corresponding x86-64 code generated by Wasmtime. The complete case is much larger.}}
    \label{motivating-example}
\end{figure}

However, even though performance debugging has long been studied~\cite{aguilera2003performance,barham2004using,linhai2014statistical}, diagnosing performance issues in \wasm runtimes remains a non-trivial task.
In general, the root causes of performance issues can be categorized into two types~\cite{jin2012understanding,zhao2022large,su2019pinpointing}.
The first type involves \textit{inefficient implementations} within applications that result in redundant computation during execution, \eg, inefficient algorithm or API usage~\cite{jin2012understanding}, unnecessary re-computations~\cite{della2015performance,song2017performance,su2019redundant}, or inefficient synchronization~\cite{ravindranath2012appinsight,weng2021argus,zhao2016non}.
Another type is \textit{suboptimal compilations}, where the compiler generates suboptimal machine code for the input program, thus causing performance issues during program execution~\cite{su2019pinpointing,theodoros2022finding}.
Existing performance debugging methods primarily target the first type; they aim to identify those redundant computations in the buggy software~\cite{xiang2023relational,linhai2014statistical,song2017performance,weng2021argus,zhao2016non}.
However, many performance issues in \wasm runtimes arise from the second type, \ie, suboptimal compilations of the input \wasm programs.
Modern \wasm runtimes typically adopt \textit{Just-In-Time compilation (JIT)}~\cite{jit}, where the input \wasm bytecode is compiled into native machine instructions before execution.
Since the JIT compiler in \wasm runtime is complicated and still immature, performance anomalies were frequently observed during the execution of the generated machine code~\cite{jiang2023revealing,Wenxuan2024Wapplique,jiang2025distinguishability}.
To diagnose such issues, it is essential to identify the specific suboptimal instruction sequences responsible for the abnormal performance.
Unfortunately, existing performance debugging methods are not well-suited for this scenario.

Identifying the exact suboptimal instruction sequences responsible for performance issues in \wasm runtimes is challenging since there are many potential candidates.
For example, as shown in Figure~\ref{motivating-example}, a \wasm program experiences abnormally long execution time on Wasmtime~\cite{wasmtime}, one of the most widely used \wasm runtimes.
The profiler \textit{perf} reported that nearly 100\% of the total time is spent executing the compiled x86-64 code of the \wasm function, \textit{func2}.
Although this function contains only 157 instructions, there are numerous potential causes of inefficiency within it, \eg, redundant zeroing (lines 56-62), redundant stores (lines 72\&77, 75-76), and excessive register spilling (lines 38, 40, 84-86) and reloads (lines 64, 66, 91-93).
As a result, manually identifying and validating which of these instruction sequences is truly responsible for the slowdown is both time-consuming and non-trivial, yet it is an essential step for further root cause analysis.

In this paper, we propose \name, a mutation-based method to pinpoint the \textit{exact} suboptimal instruction sequences responsible for performance issues (referred to as \textbf{\codes} for short) in \wasm runtimes.
\name slightly changes the original bug-inducing \wasm bytecode to obtain a functionally similar mutant, where the performance issue would \textit{not} manifest during execution.
Since the \wasm runtime would generate similar machine code for them, \name can isolate the exact \codes by comparing their corresponding machine instructions.
For example, Figure~\ref{motivating-example} illustrates the \codes identified by \name for the aforementioned case, which accurately explained the slowdown: the additional use of the register \textit{\%r10d} (lines 71-72) increases register pressure, leading to excessive register spilling and reloading, and finally causing the slowdown.
Accordingly, developers successfully inferred the root cause of this issue.
We will elaborate on this case in Section~\ref{sec_example}.

Specifically, there are two challenges to obtaining a suitable mutant for comparison.
(1) \textit{How to effectively mutate the bug-inducing \wasm program.}
Existing mutation-based fuzzing tools for \wasm~\cite{Wenxuan2024Wapplique,shangtong2024WASMaker,zhou2025lwdiff} typically apply coarse-grained mutations (\eg, at the block or function level) to improve test coverage.
However, such mutations often introduce significant functional differences from the original program, making the resulting mutants ineffective for isolating the \codes.
To this end, we design a fine-grained mutation strategy that mutates only a single \wasm instruction per mutant, thereby preserving functional similarity as much as possible.
(2) \textit{How to select functionally similar mutants that do not trigger performance issues from a series of generated mutants.}
Even with fine-grained mutations, small changes can still introduce substantial functional differences, \eg, skipping loops or introducing infinite loops.
Thus, identifying a valid mutant for comparison remains difficult.
Enlightened by the N-version programming~\cite{Algirdas1985Nversion}, which suggests that different implementations of a system can serve as a bug indicator of each other, we propose a differential selection algorithm that leverages an auxiliary runtime to filter out invalid mutants and select those most suitable for isolating the \codes.

To evaluate its effectiveness, we applied it to analyze 12 performance issues across three widely used \wasm runtimes (\ie, Wasmtime~\cite{wasmtime}, WasmEdge~\cite{wasmedge}, Wasmer~\cite{wasmer}).
Our results show that \name successfully pinpointed the exact \codes of 10 out of 12 issues (6 are previously unknown).
Furthermore, we illustrate how the \codes identified by \name assist developers in further root cause analysis.

We summarize the key contributions of this work as follows.
\begin{itemize}
    \item We conduct the first study on performance debugging in \wasm runtimes, and propose to identify suboptimal instruction sequences responsible for abnormal performance to facilitate debugging.
    \item We design and implement a mutation-based approach called \name, which aims to accurately identify suboptimal instruction sequences responsible for performance issues in \wasm runtimes. We evaluate it on 12 real-world performance issues across three widely used \wasm runtimes and verify its effectiveness compared with other approaches.
    \item We release \name as an open-source tool and have used it to successfully help diagnose six previously unknown performance issues in Wasmtime.
\end{itemize}

We make our source code and experiment data available at \url{https://github.com/BZTesting/WarpL}.



\section{Background}\label{sec_bg}

\begin{figure*}
    \centering
    \includegraphics[width=0.95\linewidth]{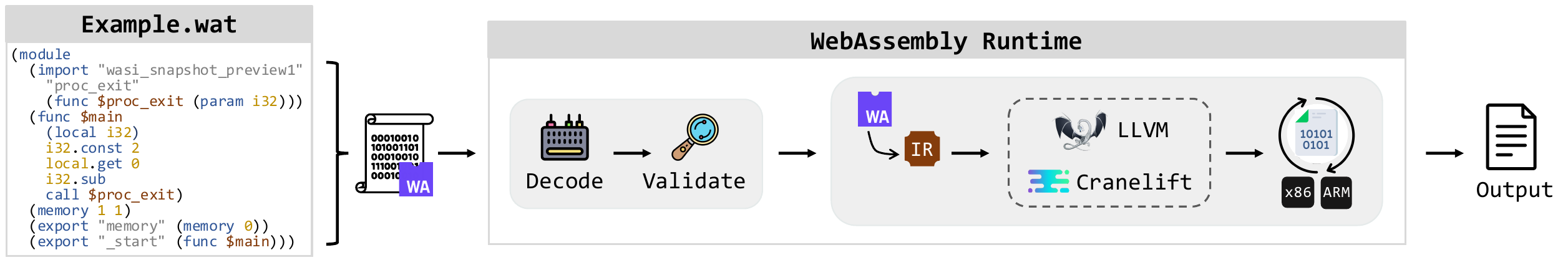}
    \caption{\textbf{An example \wasm program and the workflow of its Execution on \wasm Runtime.}}
    \label{bg_runtime}
\end{figure*}

\subsection{\wasm Runtime}
\wasm is a low-level, assembly-like programming language designed for portability, type and memory safety, high performance, and compactness~\cite{wasm}.
Due to these attributes, \wasm has been proposed to use in various scenarios, \eg, cloud computing~\cite{simon2020Faasm,philipp2022pushing,Phani2020Sledge} and trusted execution environments (TEE)~\cite{menetry2021twine, menetry2023comprehensive,wang2025waven}.
All \wasm applications run within a \wasm runtime, a sandboxed execution environment.
The general workflow of a \wasm runtime is illustrated in Figure~\ref{bg_runtime}.
Typically, \wasm runtime parses input \wasm bytecode to internal data structures based on the \wasm language specification~\cite{wasmspec}, then conducts a one-pass analysis to filter out those that fail safety checks (\eg, type check).
To improve execution performance, prevalent \wasm runtimes generally adopt \textit{Just-In-Time compilation} (JIT)~\cite{jit} to execute \wasm codes.
Specifically, the JIT compiler in \wasm runtime dynamically compiles the \wasm code into optimized executable machine code for the target architecture.
Then the runtime saves such machine codes into an executable memory region and jumps to the first machine instruction to begin execution.
When the \wasm code needs to interact with the operating system (\eg, I/O, network) during execution, it calls intrinsic functions provided by the \wasm runtime.

Prevalent \wasm runtimes utilize either LLVM~\cite{llvm} or Cranelift~\cite{cranelift} as their optimizing compilers.
LLVM is a mature compilation framework containing hundreds of optimization passes ($\sim${20M} LOC).
Cranelift, aiming to add only necessary optimizations, is a relatively young and lightweight ($\sim${0.2M} LOC) compiler and widely adopted in \wasm runtime written in Rust (\eg, Wasmer~\cite{wasmer} and Wasmtime~\cite{wasmtime}).
Therefore, compared with LLVM, Cranelift only implements some basic optimizations.
\wasm runtime compiles \wasm code function-by-function.
The compilation process usually consists of three phases:
1) In the front-end phase, \wasm runtime first compiles \wasm bytecode into the intermediate representations of the corresponding optimizing compiler (\ie, LLVM IR or Cranelift IR) in a single pass.
2) In the middle-end phase, the compiler applies a series of optimization passes to the IRs, \eg, constant folding, loop-independent code motion, and redundant-load elimination.
3) In the back-end phase, the compiler employs different instruction lowering rules, instruction selection rules, and other strategies tailored to different architectures (\eg, x86-64~\cite{x86} and aarch64~\cite{arm}) based on their characteristics to generate machine codes.
Even though the JIT speeds up the execution of \wasm programs, its complications also easily introduce more performance bugs to the runtime implementation.

\subsection{\wasm Bytecode}
\wasm has a compact binary format (\ie, \textit{bytecode}) for distribution and a text format for human-readable representation and debugging.
\wasm bytecode is organized as \textit{modules} with each module as a unit of execution.
A module contains at least one function.
Each function in a module consists of a series of \wasm \textit{instructions} which are executed in order.

\wasm instructions are stack-based~\cite{stackbased,haas2017bringing}, \ie, all operands are on an implicit \textit{operand stack} rather than in a series of registers.
During execution, \wasm instructions pop arguments from this stack and push results back onto it.
Besides immediate numbers, the data on this operand stack can be loaded from three storage sections, \ie, \textit{local variables}, \textit{global variables}, and \textit{linear memory}.
Local variables are shared within a function; global variables and linear memory are shared within a module.
During execution, data on the operand stack can also be stored back into these three storage sections.
Figure~\ref{bg_runtime} shows an example of \wasm bytecode:
\textit{i32.const 2} and \textit{local.get 0} respectively pushes the integer \textit{2} and the value of the first local variable onto the stack. \textit{i32.sub} then pops these two values from the stack as operands and pushes the subtraction result back onto the stack.

\wasm is a strongly-typed language~\cite{haas2017bringing}, which means that each executed \wasm instruction requires that the operand popped from the operand stack is of a specified type and the execution result is also a specified type.
For example, instruction \textit{i64.load} requires an i32-type value from the operand stack as the memory address to be accessed and will pop an i64-type value back to the stack.
The \wasm runtime performs a static type check to validate this; only \wasm programs that pass the type check will be executed.
Therefore, arbitrarily modifying \wasm code can easily break the type checking and be rejected by the \wasm runtime.

\section{Motivation and Challenges}\label{sec_example}

\subsection{Motivating Example}
Figure~\ref{motivating-example} presents an example, which we use to illustrate how \name helps locate the root causes of performance issues in \wasm runtimes.
The shown \wasm program experienced a long execution time on Wasmtime (over 3x slower than on WasmEdge).
The profiler \textit{perf} reported that executing the x86-64 code of the \wasm function \textit{func2} accounts for nearly 100\% of the total time.

Understanding this abnormal performance requires identifying the exact suboptimal instruction sequences responsible for the slowdown in the generated machine code.
However, this task is non-trivial, even when the code contains only 157 instructions after program reduction~\cite{wasmreduce}.
There exist numerous potential candidates within it, \eg, redundant zeroing (lines 56-62), redundant stores (lines 72\&77, 75-76), and excessive register spilling (lines 38, 40, 84-86) and reloads (lines 64, 66, 91-93).
Pinpointing the true cause among these requires substantial expertise in compiler optimizations and hardware architectures.
Moreover, manually analyzing and validating each candidate is labor-intensive and time-consuming.
Unfortunately, existing performance debugging methods~\cite{xiang2023relational,linhai2014statistical,song2017performance,weng2021argus,zhao2016non} mainly focus on identifying application-level inefficiencies and are therefore not applicable in this scenario.

\name accurately identified the \codes for this performance issue by comparing with a mutant, which differs from the original \wasm program by only one single \wasm instruction (\ie, lines 12\&13).
Both programs have nearly identical execution times on WasmEdge, but on Wasmtime, the original runs twice as slow as the mutant.
By comparing their machine code, \name found that the x86-64 code generated for the original program contains more register spilling and reload events compared to the mutant.
These events occur within a loop and directly contribute to the slowdown.
After analyzing the \wasm bytecode, we observed that Wasmtime emits two instructions (lines 71\&72) to store the value -65537 to a global variable (line 12): one to move the immediate value (\ie, -65537) into a register, and another to move the register value into memory.
In contrast, storing 65537 to the same global variable (line 13) requires only one instruction (line 73).
The extra use of a register introduces register pressure, leading to spilling and reloading, and finally causing the slowdown.
After we reported these findings to the Wasmtime developers, they confirmed that the root cause was a suboptimal instruction-lowering rule for storing signed 32-bit negative values.
The issue was fixed within two days.

\subsection{Challenges}

Obtaining such a suitable mutant involves two challenges.
(1) How to effectively mutate the bug-inducing program.
Although there exist several mutation-based generation approaches for \wasm~\cite{Wenxuan2024Wapplique,jiang2025distinguishability,shangtong2024WASMaker}, they generally mutate the \wasm bytecode at a coarse granularity, \eg, blocks containing dozens of \wasm instructions~\cite{Wenxuan2024Wapplique,jiang2025distinguishability} or entire functions~\cite{shangtong2024WASMaker,zhou2025lwdiff}, to gain higher test coverage.
These approaches tend to generate mutants that are functionally different from the original \wasm program, and thus are inapplicable in this scenario.
To isolate the \codes, we need a fine-grained mutation strategy that ensures each mutant differs from the original program only slightly. (\textbf{Section \ref{app_1}})
(2) How to select, from the generated mutants, one that is functionally similar to the original program and does not trigger the performance issue.
Even minor mutations can lead to significant semantic differences.
Worse still, such invalid mutants usually exhibit performance behavior changes (\eg, shorter execution time) on the buggy \wasm runtime, appearing as if the performance issue is not triggered.
This makes it difficult to distinguish a suitable mutant from invalid ones.
Therefore, we need to design a reliable selection algorithm that identifies, among a large number of generated mutants, those that remain functionally equivalent while avoiding the performance anomaly. (\textbf{Section \ref{app_2}})

\section{Approach}\label{sec_app}

\begin{figure*}
    \centering
    \includegraphics[width=0.9\linewidth]{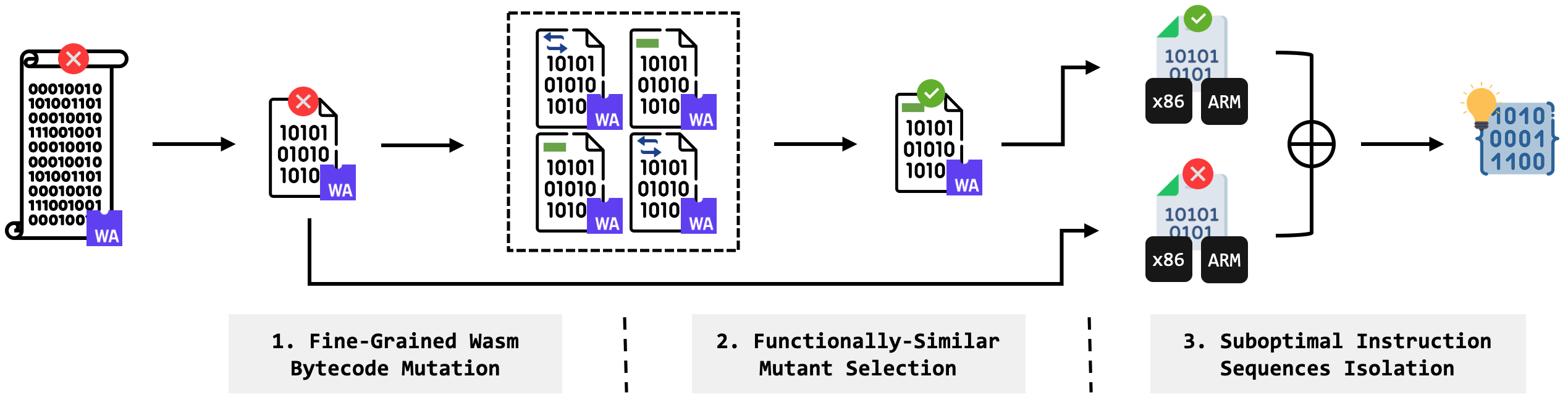}
    \caption{\textbf{The workflow of \name.}}
    \label{approach}
\end{figure*}

\subsection{Overview}
Figure~\ref{approach} presents the overall workflow of \name.
It consists of three main steps: (1) fine-grained \wasm bytecode mutation, (2) functionally similar mutant selection, and (3) suboptimal instruction sequences isolation.
Since reported bug-inducing \wasm programs are often long (thousands of lines), we first apply program reduction to remove irrelevant code and improve mutation efficiency.
\name then generates a set of mutants using the fine-grained mutation strategy (\textbf{Section \ref{app_1}}), ensuring that each mutant differs from the original program by exactly one \wasm operation.
Next, \name picks out the functionally similar mutants that do not trigger performance issues by a differential selection algorithm (\textbf{Section \ref{app_2}}).
\name then extracts the machine code generated by the buggy \wasm runtime for both the reduced program and the selected mutant, and compares them to isolate the \codes (\textbf{Section \ref{app_3}}).
Finally, \name produces a visual report highlighting these differences, as shown in Figure~\ref{motivating-example}.
We describe the design and implementation details of \name in the following sections.

\subsection{Fine-Grained \wasm Bytecode Mutation}\label{app_1}

In this subsection, we introduce our fine-grained \wasm bytecode mutation strategy and explain how it is used to generate mutants.

The core idea of the fine-grained mutation is to mutate exactly one \wasm instruction per mutation, while avoiding inserting new \wasm instructions whenever possible.
However, since \wasm is a strongly-typed language, randomly deleting or modifying an instruction can easily lead to type errors, causing the mutated program to fail runtime validation.
To ensure that the generated programs are valid, we design a \textit{type-aware mutation strategy}.
Specifically, we categorize \wasm instructions into two groups\footnote{We exclude all \wasm control instructions~\cite{wasm-ctrl-instr} to avoid directly altering the control flow.}:
\begin{itemize}[leftmargin=*]
    \item \textbf{Operand instructions}, which push data onto the implicit operand stack. The data includes immediate constants, local or global variables, and linear memory values.
    \item \textbf{Operator instructions}, which consume operands from the stack to perform computations. Some of them also push results back onto the stack.
\end{itemize}

We handle mutations for these two categories separately.

\textbf{Operand instruction substitution.}
We replace an operand instruction with another of the same data type.
\wasm supports four numeric types: \textit{i32}, \textit{i64}, \textit{f32}, and \textit{f64}.
If the target is an immediate constant instruction, we either modify its constant value (keeping the data type) or replace it with a local/global variable load instruction of the same data type.
If the target is an instruction that loads a value from a local/global variable or linear memory, we replace it with an immediate constant instruction of the same data type.

\textbf{Operator instruction substitution.}
We replace an operator instruction with another that has the same \textit{operator type}.
In \wasm, an operator type defines how an instruction interacts with the implicit operand stack.
It is denoted as \( \textit{\textit{optype}} := [\text{\textit{type}}^*] \rightarrow [\text{\textit{type}}^*] \), where \textit{type} denotes a \wasm data type.
For example, \textit{i64.eq} has the operator type \( [\text{\textit{i64}}, \text{\textit{i64}}] \rightarrow [\text{\textit{i32}}] \), meaning it pops two \textit{i64} values from stack and pushes back one \textit{i32} result.
We group operator instructions by their operator types according to the \wasm specification~\cite{wasmspec}, and only allow substitutions within the same group during mutation.

\textbf{Operator instruction deletion.}
We also allow deleting certain operator instructions.
To preserve type validity, we delete both the target operator instruction and its corresponding operand-producing instructions, and insert suitable immediate constant instructions to restore the expected stack state.
To minimize functional changes, we only delete operator instructions whose operands are all produced by operand instructions, not other operator instructions.

Our mutation rules are summarized in Table~\ref{tab:mutator}.
During mutation, \name traverses every \wasm instruction in the bug-inducing program (excluding control instructions) and applies each mutation rule to generate mutants.

\begin{table}
    \centering
    \small
    \caption{Fine-Grained and Type-Aware \wasm Mutators}
    \begin{tabular}{p{1.5cm}|p{6.3cm}}
        \toprule
            \textbf{Instruction Categories} & \textbf{Rules} \\
        \midrule
            \multirow{1}{*}{
                \parbox[c][0.3cm][c]{1.5cm}{\raggedright Operand Instructions}
            }
            &
            $
            \begin{array}{l}
                \textbf{Rule1 - Operand\ Instruction\ Substitution:} \\
                \llbracket \mathit{t.const\ imm} \rrbracket \rightarrow \llbracket \mathit{t.const\ imm'} \rrbracket \\
                \quad |\ \llbracket \mathit{local.get\ k} \rrbracket,\ \mathrm{where}\ \mathit{type(local[k]) = t} \\
                \quad |\ \llbracket \mathit{global.get\ k} \rrbracket,\ \mathrm{where}\ \mathit{type(global[k]) = t} \\
                \llbracket \mathit{local.get\ k} \rrbracket \rightarrow \\
                \quad \llbracket \mathit{t.const\ imm} \rrbracket,\ \mathrm{where}\ \mathit{type(local[k]) = t} \\
                \llbracket \mathit{global.get\ k} \rrbracket \rightarrow \\
                \quad \llbracket \mathit{t.const\ imm} \rrbracket,\ \mathrm{where}\ \mathit{type(global[k]) = t} \\
                \llbracket \mathit{i32.const\ addr; t.load} \rrbracket \rightarrow \llbracket \mathit{t.const\ imm} \rrbracket
            \end{array}
            $\\
        \midrule
             \multirow{2}{*}{
                \parbox[c][1.8cm][c]{1.5cm}{\raggedright Operator Instructions}
            }
            &
            $
            \begin{array}{l}
                \textbf{Rule2 - Operator\ Instruction\ Substitution:} \\
                \llbracket \mathit{t.op} \rrbracket \rightarrow \llbracket \mathit{t.op'} \rrbracket,\ \mathrm{where}\ \mathit{type(t.op') = type(t.op)} \\ [3pt]
            \end{array}
            $ \\
            \cline{2-2}
            &
            $
            \begin{array}{l}
                \noalign{\vspace{2pt}}
                \textbf{Rule3 - Operator\ Instruction\ Deletion:} \\
                \llbracket \mathit{(t.const\ imm\ |\ local.get\ k\ |\ global.get\ k} \\
                |\ \mathit{(i32.const\ addr; t.load))^*; t.op} \rrbracket_{e_1} \rightarrow \\
                \quad \llbracket \mathit{(t.const\ imm)^*} \rrbracket_{e_2}, \ \mathrm{where}\ \mathit{type(e2) = type(e1)}
            \end{array}
            $ \\
        \bottomrule
    \end{tabular}
    \label{tab:mutator}
\end{table}

\subsection{Functionally-Similar Mutant Selection}\label{app_2}

In this subsection, we describe how we select a functionally similar mutant that does not trigger the performance issue from a set of generated mutants.

Although our fine-grained mutation strategy mutates only one \wasm instruction per mutant, such changes can still introduce substantial functional differences.
For example, a mutation may alter a loop condition, causing the loop to be skipped and thereby changing program behavior.
These invalid mutants often exhibit large performance differences compared to the original program when executed on the buggy \wasm runtime.
However, since there is no absolute metric to indicate performance issues, a mutant with a large performance difference from the original program is typically regarded as not triggering the performance issue.
As a result, it is difficult to distinguish valid, functionally similar mutants from invalid mutants based only on performance behaviors on the buggy \wasm runtime.
To solve this problem, we leverage an auxiliary runtime\footnote{We refer to it as \textit{oracle runtime} in the rest of the paper.} to filter out invalid mutants.
Specifically, we compare the performance of the original program and each mutant on both the buggy and oracle \wasm runtimes.
If the performance difference is small on the oracle runtime but large on the buggy runtime, it suggests that the mutant is functionally similar to the original program but no longer triggers the performance issue.

\begin{algorithm}[t]
    \small

    \caption{Scoring Algorithm For Mutants}\label{algo_scoring}
    \SetKwInOut{Input}{Input}
    \SetKwInOut{Output}{Output}
    \SetKwFunction{collectPerfMetricInBuggyRt}{collectPerfMetricInBuggyRt}
    \SetKwFunction{collectExeTimeInOracleRt}{collectExeTimeInOracleRt} 

        \Input{$originalProgram$, $mutantSet$}
        \Output{$scoreSet$}

        $oPerfMetricInBuggyRt \leftarrow$ \collectPerfMetricInBuggyRt{$originalProgram$}\;
        $oExeTimeInOracleRt \leftarrow$ \collectExeTimeInOracleRt{$originalProgram$}\;
        $scoreSet \leftarrow$ []\;
        
        \ForEach{$variant \in mutantSet$} {
            $vPerfMetricInBuggyRt \leftarrow$ \collectPerfMetricInBuggyRt{$variant$}\;
            $vExeTimeInOracleRt \leftarrow$ \collectExeTimeInOracleRt{$variant$}\;
            $perfDiffRatio \leftarrow \frac{oPerfMetricInBuggyRt}{vPerfMetricInBuggyRt}$\;
            $funcSimRatio \leftarrow \frac{oExeTimeInOracleRt}{vExeTimeInOracleRt}$\;
            \lIf{$perfDiffRatio > 1$}{$perfDiffScore \leftarrow 1 - \exp(1-perfDiffRatio) $}
            \lElse{$perfDiffScore \leftarrow 1 - perfDiffRatio^2 $}

            \lIf{$funcSimRatio > 1$}{$funcSimScore \leftarrow \exp(1-funcSimRatio) $}
            \lElse{$funcSimScore \leftarrow funcSimRatio^2 $}

            $mutantScore \leftarrow \alpha*perfDiffScore + \beta*funcSimScore $\;
            $scoreSet \leftarrow scoreSet \cup [mutantScore] $\;
        }

\end{algorithm}

To implement this idea, we collect the performance data for the original program and each mutant on both buggy and oracle runtimes, and compute a score for each mutant using Algorithm~\ref{algo_scoring}.
The final score is a weighted sum of two components: \texttt{perfDiffScore} and \texttt{funcSimScore} (line 13).
\texttt{perfDiffScore} (line 7, 9, 10) measures the performance differences between the mutant and the original program on the buggy runtime; a higher value (closer to 1) indicates a larger performance difference.
\texttt{funcSimScore} (line 8, 11, 12) quantifies the similarity in execution time between the mutant and the original program on the oracle runtime; a higher value (closer to 1) implies greater functional similarity.
After scoring all mutants, we select the one with the highest score for isolating the suboptimal runtime events.

\subsection{Suboptimal Instruction Sequences Isolation}\label{app_3}

In this subsection, we illustrate how \name isolates the suboptimal instruction sequences responsible for the performance issue in the runtime-generated machine code.

Most modern \wasm runtimes adopt just-in-time (JIT) compilation to execute \wasm functions, and typically provide interfaces for dumping the generated machine code.
For runtimes that lack such support (\eg, WasmEdge~\cite{wasmedge}), it is straightforward to insert additional code to dump the compiled machine code after JIT compilation.
Once the machine code for both the original program and the selected mutant is obtained from the buggy \wasm runtime, \name compares them at the assembly level.
Since the register allocation may vary significantly between these two machine codes, we compare only their instruction opcodes using the Longest Common Subsequence (LCS) algorithm~\cite{LCS}.
Besides, we compare the total number of instructions and the address of the first instruction for each \wasm function.
Finally, \name generates a visual report summarizing these differences, as illustrated in Figure~\ref{motivating-example}.

\section{Evaluation}\label{sec_eval}

%

   

\begin{table*}[t]
\centering
\small
\caption{\textbf{\name's result on 12 performance issues across three \wasm runtimes.}
"Succ?" shows whether \name \textit{successfully} identifies the exact \codes.
"Rank" shows the ranking of the mutants that helps debugging successfully.
"Ratio(B)" shows the ratio of the execution time of the original program to that of the selected mutant in \textbf{B}uggy runtime.
"Ratio(O)" shows the ratio of the execution time of the original program to that of the selected mutant in \textbf{O}racle runtime.
"\#MI(R)" shows the \textit{machine instructions} number of the \textbf{Reduced} \wasm program.
"\#MI(I)" shows the number of \textit{machine instructions} \textbf{Identifed} by \name.
}
\begin{tabular}{l|cc|c|ccc|ccc}
 \toprule
 \textbf{Issue}\textsuperscript{*} & Buggy Runtime  & Oracle Runtime & Succ? & Rank & Ratio(B) & Ratio(O) & \#MI(R) & \#MI(I)  \\
 \midrule
 \#7085 & Wasmtime:f10d66 & WasmEdge:dc2f26 & Yes & 1st & 7.77 & 1.01 & 39 & 2  \\ 
 \#8573 & Wasmtime:d0cf46 & WasmEdge:dc2f26 & Yes & 1st & 2.00 & 1.00 & 96 & 1 \\ 
 \#8571 & Wasmtime:7f7064 & WasmEdge:dc2f26 & Yes & 1st & 4.58 & 1.00 & 38 & 4 \\ 
 \#8706 & Wasmtime:7f7064 & WasmEdge:dc2f26 & Yes & 4th & 1.84 & 1.00 & 177 & 25  \\ 
 \#9590 & Wasmtime:c8b136 & WasmEdge:eaf8d0 & Yes & 3rd & 6.19 & 0.99 & 51 & 11 \\ 
 \midrule
 \#7246 & Wasmtime:f10d66 & WasmEdge:dc2f26 & Yes & 1st & 59.94 & 0.99 & 29 & 0 \\ 
 \#7731 & Wasmtime:6613ac & WasmEdge:dc2f26 & Yes & 1st & 6.81 & 1.00 & 35 & 2 \\ 
 \#3784 & Wasmer:148021 & WasmEdge:dc2f26 & Yes & 3rd & 28.25 & 1.00 & 62 & 3   \\  
 \#2442 & WasmEdge:dc2f26 & Wasmtime:7f7064 & Yes & 1st & 1.51 & 0.99 & 135 & 35  \\ 
 \#2963 & WasmEdge:gabc43 & Wasmtime:d51b5a & Yes & 2nd & 1.68 & 1.12 & 37 & 4 \\ 
 \#7973 & Wasmtime:6613acd & WasmEdge:dc2f26 & No & - & - & - & - & - \\
 \#7745 & Wasmtime:6613acd & WasmEdge:dc2f26 & No & - & - & - & - & - \\
 
 \bottomrule
\end{tabular}
\label{tab_issues}
\end{table*}

\textit{\name} aims to identify the \codes, \ie, suboptimal instruction sequences responsible for performance issues in \wasm runtimes, to facilitate further root cause analysis.
We evaluate its effectiveness by answering the following research questions (RQs):
\begin{itemize}
    \item \textbf{RQ1:} Can \name identify the exact \codes responsible for performance issues in \wasm runtimes?
    \item \textbf{RQ2:} How does the \codes identified by \name assist in locating the root causes of performance issues?
    \item \textbf{RQ3:} How effective are the fine-grained \wasm mutation strategy and the differential selection algorithm in identifying the \codes?
\end{itemize}

\textbf{Implementations.}
Since abnormally long execution time is one of the most common symptoms of performance issues in \wasm runtimes, our experiments focus on this type of issue.
Accordingly, we use the program execution time as the performance metric.
We implement our approach as a tool in approximately 400 lines of C++ code using Binaryen's LibTooling~\cite{binaryen} for \wasm bytecode mutation, and around 500 lines of Python code and shell scripts for functionally-similar mutant selection, suboptimal runtime events isolation, and visual report generation.
Besides, we added a small patch to WasmEdge~\cite{wasmedge} to enable dumping of generated machine code.


\textbf{Target \wasm Runtimes and Evaluated Performance Issues.}
We evaluate \name on 12 performance issues across the top three \wasm runtimes by GitHub stars: Wasmtime~\cite{wasmtime}, Wasmer~\cite{wasmer}, and WasmEdge~\cite{wasmedge}.
These runtimes are complicated and comprise over 455K, 252K, and 108K lines of code, respectively.
Among the 12 issues, 5 were newly discovered in Wasmtime using WarpDiff~\cite{jiang2023revealing} and Wapplique~\cite{Wenxuan2024Wapplique} (\ie, \#7085, \#8573, \#8571, \#8706, \#9590).
We selected Wasmtime as the detection target for two reasons:
1) It is actively maintained, with daily updates.
2) It always integrates the latest version of the Cranelift compiler.
Moreover, both Wasmtime and Cranelift are developed by the same team, which facilitates timely communication with developers.
In contrast, Wasmer relies on an outdated Cranelift API, and WasmEdge uses LLVM as its backend.
The remaining 7 issues were collected from the issue trackers of the three runtimes.
We searched using the keyword ``performance'' and selected posts that included terms ``poor'', ``degradation'', or ``bad'' and provided input \wasm programs that trigger the issues.
We excluded those that could not be reliably reproduced.
In total, we obtained 12 performance issues, as summarized in Table~\ref{tab_issues}.

Since the bug-inducing \wasm programs are all long, before debugging, we first applied \textit{wasm-reduce}~\cite{wasmreduce} to remove irrelevant code before mutation.
Specifically, we adopted \textit{wasm-reduce} to prune the \wasm program greedily, and integrated a validation mechanism to ensure that the reduced program still preserved the original performance issue.
The validation checks whether (1) the execution time of the reduced program on the buggy runtime is comparable to the original, and (2) the relative performance difference between buggy and oracle runtimes remains similar.

\textbf{Experimental environment.}
We conducted our experiments on two systems: one with an Intel(R) Core(TM) i5-13500 running Ubuntu 22.04, and another with an Intel(R) Core(TM) i5-9500T CPU @ 2.20GHz running Ubuntu 20.04.
All experiments were independently run on both systems, and the results were consistent across them, except for Issue \#8573, which could only be reproduced on the i5-9500T platform.

\begin{table*}
\centering
\small
\caption{\textbf{Summary of root causes for previously-unknown Issues.}}
\begin{tabular}{l|c|c}
    \toprule
    \textbf{Issue} & Type & Summary of the root causes  \\
    \midrule
    \#7085 & ISA Specific & Overlooks the register dependencies in \textit{vcvtsi2sdl} instruction.                 \\
    \#8573 & ISA Specific &  Sets inappropriate function alignment settings.                              \\
    \#8571 & Missed Optimization & Suboptimal side-effect analysis when eliminating dead division.  \\
    \#8706 & ISA Specific & Suboptimal instruction lowering rules for negative 32-bit values.   \\
    \#9590 & Missed Optimization & Suboptimal dead code elimination.  \\
    \#7246 & Wrong Optimization & Incorrectly sinks a subnormal floating-point multiplication into the loop.                  \\
    \bottomrule
\end{tabular}
\label{tab_root_cause_summaries}
\end{table*}

\subsection{Effectiveness for Identifying \codes}\label{eval_rq1}

Table~\ref{tab_issues} summarizes the effectiveness of \name in identifying the \codes for twelve performance issues in \wasm runtimes.
Overall, \name successfully located the exact \codes for ten issues.
The mutants that led to successful isolation were consistently ranked among the top four candidates, with six of them ranked highest.
For the five previously unknown issues, the \codes identified by \name provided explanations of the root causes that were confirmed by the developers.
We first submitted the \codes isolated by the top-ranked mutant.
If no root cause was confirmed within one day, we submitted the result from the next highest-ranked mutant, and repeated this process until the issue was resolved.
For the seven closed issues, we considered \name successful if its identified \codes captured the root cause documented in the issue tracker.
Notably, for Issue \#7246, we found that the originally reported root cause was incomplete.
\name helped uncover the actual underlying cause, which is subtle and hard to observe directly.
We will elaborate on it in Section~\ref{eval_rq2}.

\name was unable to identify the \codes for Wasmtime-\#7973 and Wasmtime-\#7745, as these issues were not caused by suboptimal compilation, but rather by inefficient I/O implementations within Wasmtime.
In both cases, the profiler \textit{perf} reported that the hot functions resided in external library code used by Wasmtime, not in the JIT-generated code.
Most existing performance debugging techniques~\cite{xiang2023relational,linhai2014statistical,song2017performance,weng2021argus} are designed to analyze such cases and perform well when the inefficiency lies at the algorithm level.
However, they are generally ineffective in identifying suboptimal instruction sequences in runtime-generated code, while \name fills this gap.
As shown in Table~\ref{tab_issues}, \name accurately narrowed down the abnormal performance to a small set of instructions.
These instructions significantly assisted developers in further root cause analysis, as discussed in Section~\ref{eval_rq2}.
This effectiveness is attributed to \name's fine-grained \wasm bytecode mutation strategy (Section~\ref{app_1}) and functionally similar mutant selection algorithm (Section~\ref{app_2}).
We further evaluate the contribution of these components by comparing them with alternative methods in Section~\ref{eval_rq3}.

\subsection{Helpfulness for Root Cause Analysis}\label{eval_rq2}

In this section, we present comprehensive case studies to illustrate how \name facilitates root cause analysis, particularly for previously unknown issues.
After we shared the results output by \name, \textbf{the developers successfully committed patches to fix Issues \#7085, \#8573, and \#8706, linked Issue \#8571 to an ongoing pull request, and confirmed the root causes of Issues \#9590 and \#7246}.
The root causes for these issues are summarized in Table~\ref{tab_root_cause_summaries}.
For all six issues, we engaged in multiple rounds of discussions with the developers.
We found that the suboptimal instructions identified by \name not only explained the direct causes of the slowdowns but also, when combined with analysis of \wasm bytecode differences, provided critical clues for inferring the underlying root causes within the runtime implementations.
During our interactions, developers offered positive feedback and described the findings as both ``interesting'' and ``helpful''.
Below, we summarize these discussions and present them as case studies.
Issue \#8706 has been introduced earlier in Section~\ref{motivating-example}.

\begin{figure}[t]
    \centering
    \small
    \includegraphics[width=0.9\linewidth]{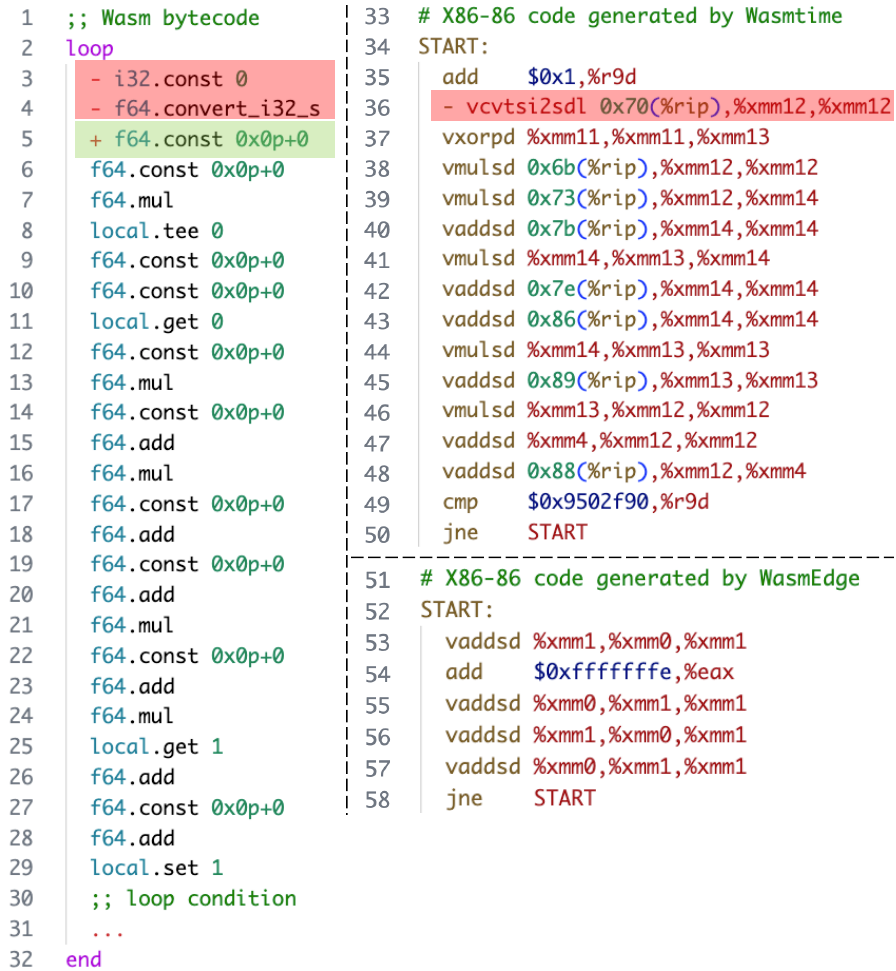}
    \caption{Code related to Issue \#7085. The left side shows the differences between the bug-inducing program (red) and the mutated program (green). The right side shows their differences in the machine code. Same as the other issues.} 
    \label{fig_issue1}
\end{figure}

\textbf{Issue \#7085: }\label{issue001}
Figure~\ref{fig_issue1} shows the results generated by \name for Issue \#7085.
The root cause of this slowdown is that Wasmtime mishandled register dependencies in the \textit{vcvtsi2sdl} instruction when compiling \wasm instruction \textit{f64.convert\_i32\_s}.
\name highlights the direct cause for the slowdown in line 36.
Instruction \textit{vcvtsi2sdl 0x70(\%rip),\%xmm12,\%xmm12} (line 36) converts a 64-bit integer at the address \textit{rip+0x70} to a 64-bit float number and sets it in the lower 64 bits of the \textit{xmm12} register.
It also copies the upper 64 bits of the \textit{xmm12} back into \textit{xmm12}.
Since subsequent instructions within the loop (lines 46, 47) also write to the \textit{xmm12}, this creates a register dependency across iterations.
Such dependencies hinder the out-of-order execution capabilities of modern CPUs, leading to reduced performance~\cite{firefox}.
We manually wrote a microbenchmark in assembly to validate this and included it in our report.

The associated \wasm bytecode identified by \name (lines 3–4) helps infer the root cause further. 
This \wasm code converts a 32-bit integer 0 to a 64-bit float.
Wasmtime uses the x86-64 instruction \textit{vcvtsi2sdl} to execute the \wasm instruction \textit{f64.convert\_i32\_s}.
Since the \textit{vcvtsi2sdl} is a 128-bit instruction while \textit{f64.convert\_i32\_s} uses only the lower 64 bits, Wasmtime ignores that the destination register also depends on the second source register during the instruction lowering.
Therefore, the unexpected register dependency (\ie, \textit{xmm12}) causes this slowdown.

Before we submitted our findings, the developers had already investigated several potential causes.
They suspected that the slowdown was due to missing optimizations in Wasmtime, particularly because the x86-64 code generated by Wasmtime is twice as long as that from WasmEdge (see Figure~\ref{fig_issue1}). 
They first assumed that the inefficient memory-based constant loading (\ie, lines 38–40) was the cause of the slowdown.
However, after modifying Cranelift’s constant handling mechanism, they found that the performance degradation persisted.
They also attempted to hoist recalculations (\eg, lines 38-40) out of the loop, but this strategy also failed to resolve the issue.
Ultimately, one developer remarked, \textit{``... and I couldn't explain what was happening. This appears to be some sort of micro-architectural cliff, but we couldn't figure out how to actually see the cliff.''}.

\begin{figure}[t]
    \centering
    \small
    \includegraphics[width=0.85\linewidth]{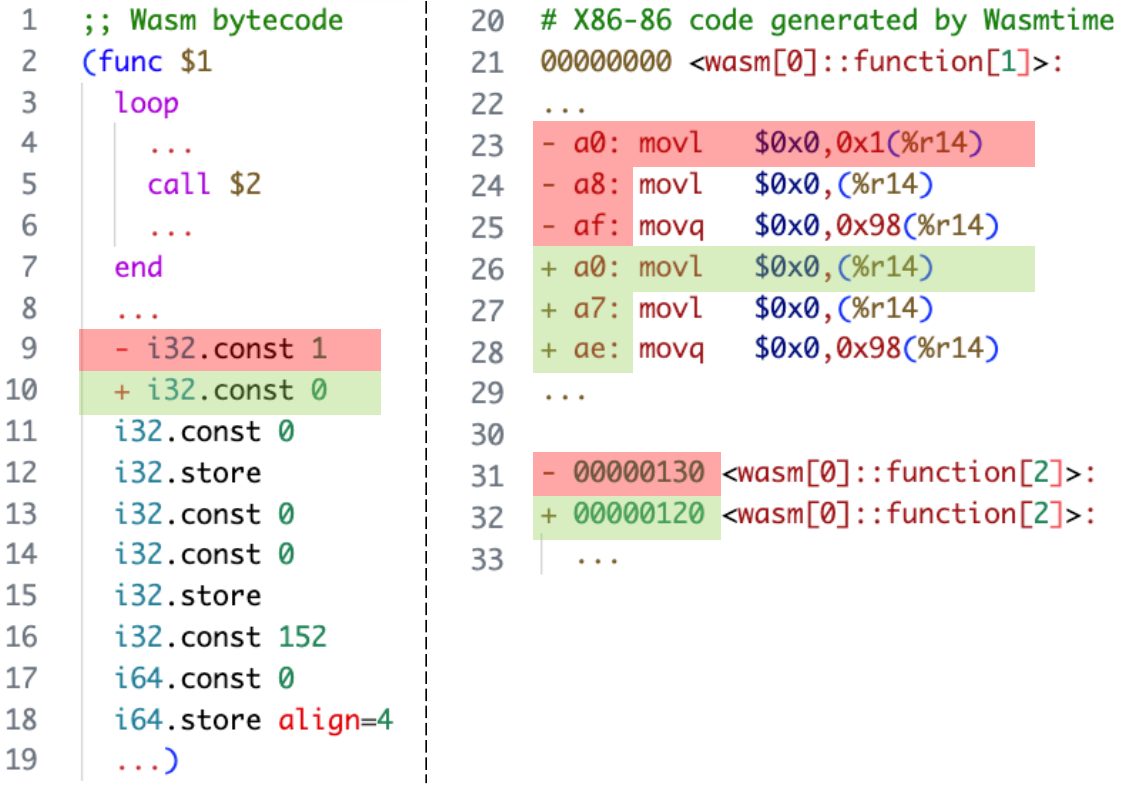}
    \caption{Code related to Issue \#8573.}
    \label{fig_issue2}
\end{figure}

\textbf{Issue \#8573: }
Our result for Issue \#8573 is shown in Figure~\ref{fig_issue2}.
The slowdown was caused by inappropriate function alignment during code generation in Wasmtime.
\name found that the x86-64 instruction sequences of the original and mutated programs were nearly identical.
This prompted developers to notice a subtle yet critical difference: the starting address of \textit{func2}.
In the original program, \textit{func2} begins at address \textit{0x130}, which is 16-byte aligned.
However, modern CPU frontends typically fetch 32-byte or 64-byte aligned chunks from the instruction cache.
When a function starts mid-cache-line, additional fetches may be required, which incurs overhead.
Since \textit{func2} is invoked repeatedly in a loop, this overhead accumulates and results in a noticeable slowdown.

We now illustrate why \name can guide developers to uncover the root cause.
In the original program, the \wasm instruction \textit{i32.const 1} (line 9) is compiled into the assembly \textit{movl \$0x0,0x1(\%r14)}, whose machine code is \textit{41 c7 46 01 00 00 00 01} (occupying 8 bytes, from address \textit{a0} to \textit{a7} in line 23).
As a result, the machine code for \textit{func1} spans from \textit{0x0} to \textit{0x120}, and with Wasmtime's 16-byte alignment for x86-64, \textit{func2} begins at \textit{0x130}.
In contrast, in the mutated program, the \wasm instruction \textit{i32.const 0} (line 10) is compiled into the assembly \textit{movl \$0x0,(\%r14)}, whose machine code is \textit{41 c7 06 00 00 00 00} (7 bytes, from address \textit{a0} to \textit{a6} in line 26).
Consequently, \textit{func1} ends at \textit{0x11f}, and \textit{func2} starts at \textit{0x120}.
Developers noticed that \textit{0x120} is 32-byte aligned, whereas \textit{0x130} is only 16-byte aligned, which inspired developers to infer function alignment as the root cause.
Without \name, this issue would have been difficult to diagnose, as developers typically focus on inefficient instruction sequences when debugging rather than on subtle differences in code layout within memory.

\begin{figure}[t]
    \centering
    \small
    \includegraphics[width=0.85\linewidth]{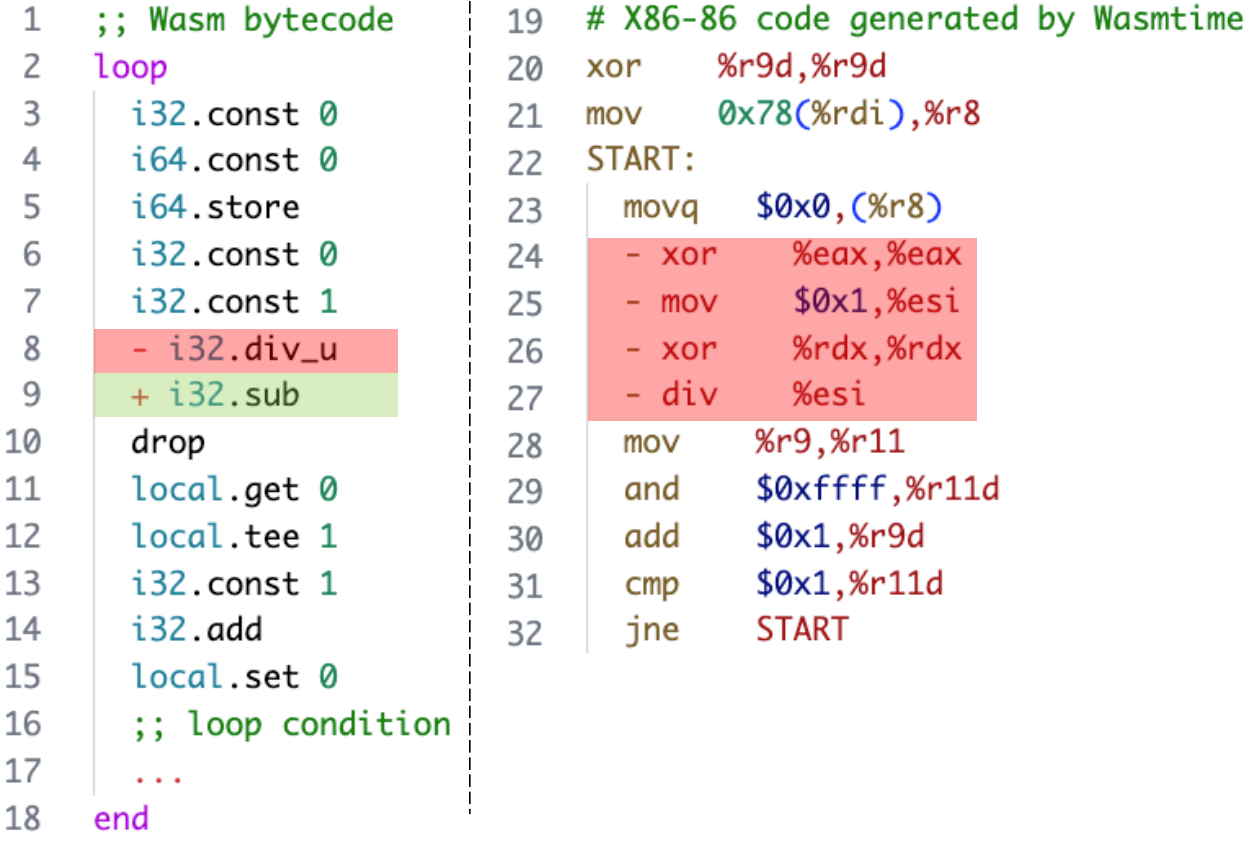}
    \caption{Code related to Issue \#8571.}
    \label{fig_issue3}
\end{figure}

\textbf{Issue \#8571: }
We present the result generated by \name for Issue \#8571 in Figure~\ref{fig_issue3}.
The reason for this slowdown is that Wasmtime fails to eliminate the dead division during compilation.
\name highlighted the direct cause for the slowdown in lines 24-27.
Since division (line 27) is an expensive operation on many CPUs, the costs are much magnified within a loop, ultimately causing this performance issue.

The root cause was further clarified by analyzing the \wasm bytecode differences in lines 8–9. 
In the original program, the division (line 8) involves a constant dividend of \textit{0} (line 6) and a divisor of \textit{1} (line 7); its result is unused (line 10), making it dead code.
In contrast, the mutated program replaces the division with a subtraction (line 9), which is successfully optimized away by Wasmtime.
This finding quickly inspired developers to infer the root cause:
Wasmtime thinks that the division might have side effects, \ie, if the divisor is 0, then the runtime should raise an error.
Although the divisor is a known constant (\texttt{1}), existing side-effect analysis in Wasmtime does not consider constant values, and thus fails to eliminate the division.
Without \name, identifying the specific optimization missed by Wasmtime would be considerably more time-consuming, especially since the WasmEdge-compiled x86-64 code shows that the entire loop can be eliminated.

\begin{figure}[t]
    \centering
    \small
    \includegraphics[width=0.9\linewidth]{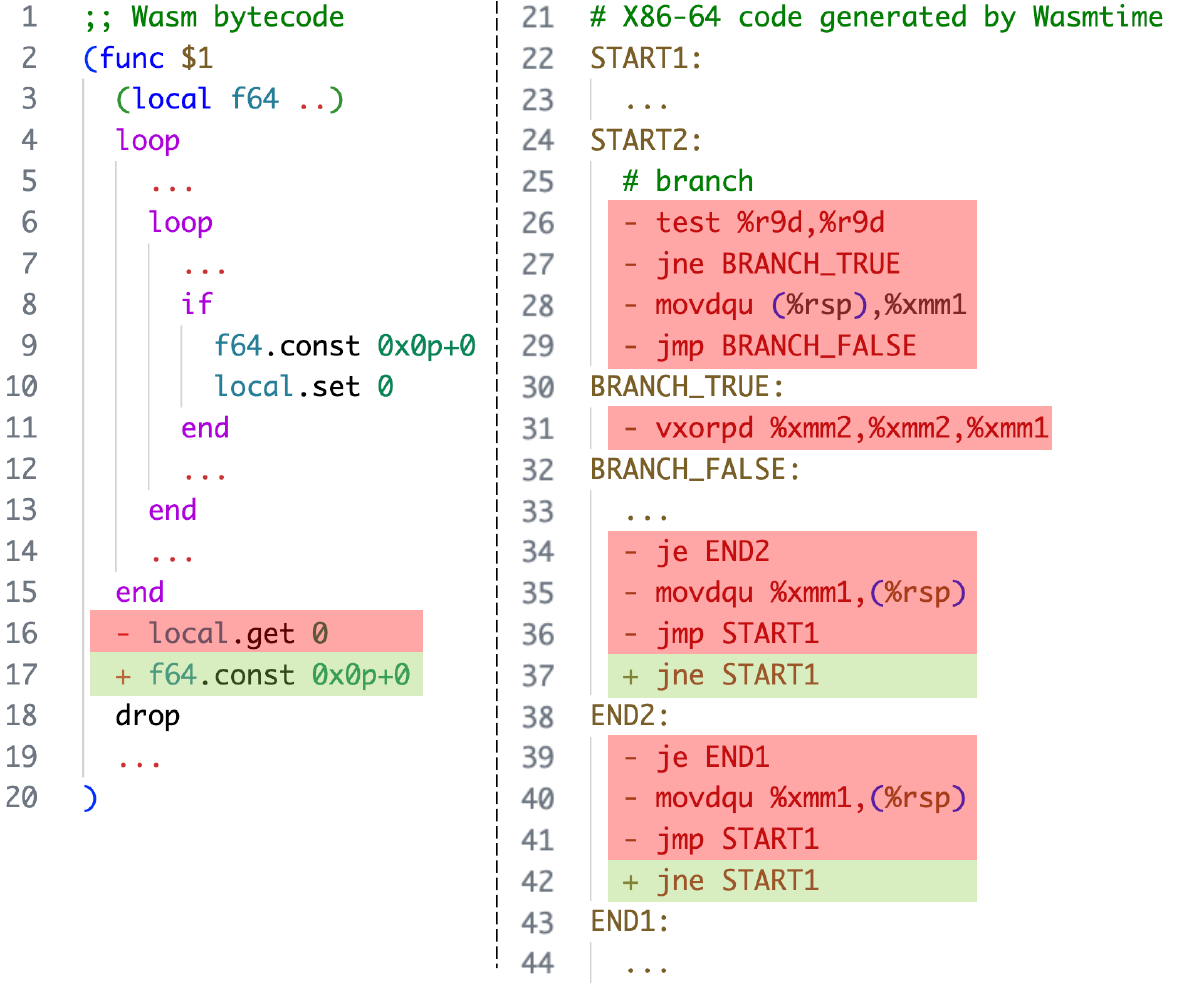}
    \caption{Code related to Issue \#9590.}
    \label{fig_issue5}
\end{figure}

\textbf{Issue \#9590: }
The result generated by \name for Issue\#9590 is presented in Figure~\ref{fig_issue5}.
The slowdown is due to suboptimal dead code elimination in Wasmtime.
As highlighted by \name, the x86-64 code of the original program contains more register spilling (lines 35, 40), register reloads (line 28), and branch instructions compared to the mutated program.
These additional instructions occur within a loop, directly resulting in the observed slowdown.

Further analysis of the \wasm bytecode differences (lines 16–17) reveals the underlying cause.
The local variable \textit{local[0]} is conditionally modified (lines 8–11) within a nested loop and is only accessed once outside the loop (line 16).
At the beginning of every loop iteration, the value of \textit{local[0]} is stored in memory at address \textit{\%rsp}, while a temporary copy is kept in the register \textit{\%xmm1} (line 28).
If the loop condition holds, \textit{local[0]} is updated (line 31), and at the end of the iteration, the updated value in \textit{\%xmm1} is written back to \textit{\%rsp}.
This frequent update incurs repeated register spills and reloads.
However, we observed that the instruction accessing \textit{local[0]} (line 16) is immediately followed by a \textit{drop} operation (line 18), indicating that the value is never actually used.
This suggests that Wasmtime's dead code analysis for local variable access is suboptimal: since \textit{local[0]} is not semantically used, all related computations inside the loop should have been eliminated.
We submitted our findings to the developers, and they confirmed the root cause.
Without the highlighted code differences, pinpointing the issue from such long instruction sequences would have been extremely difficult.

\begin{figure}[t]
    \centering
    \small
    \includegraphics[width=0.85\linewidth]{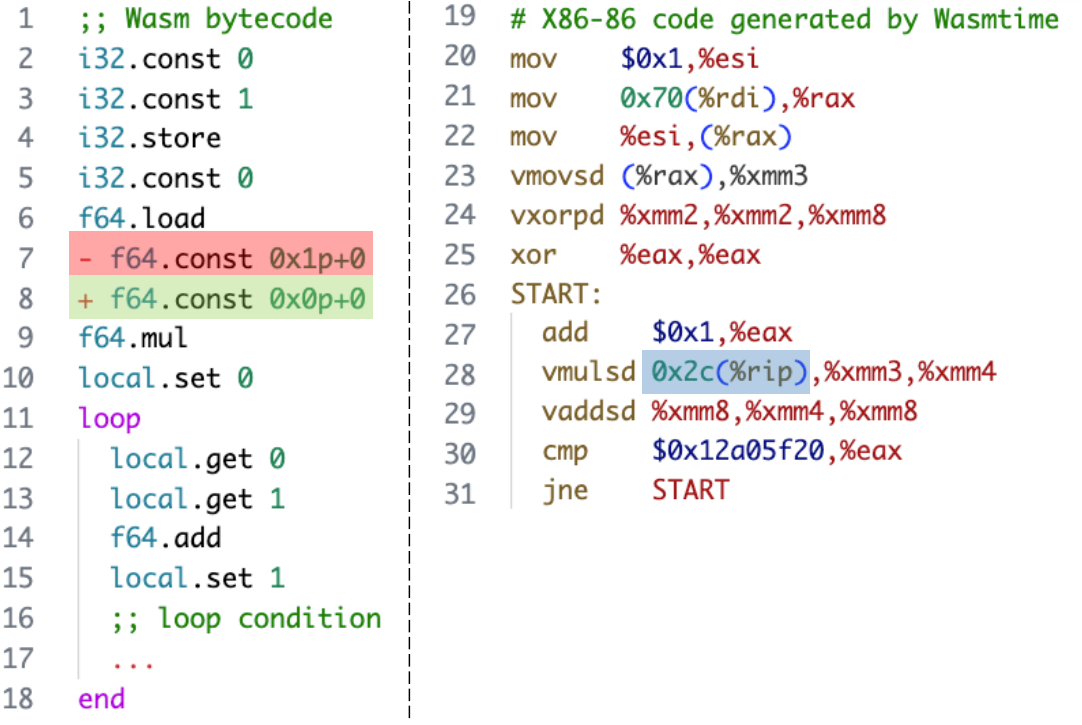}
    \caption{Code related to Issue \#7246. The highlighted constant in \wasm code is stored in memory address \textit{rip+0x2c}.}
    \label{fig_issue6}
\end{figure}

\textbf{Issue \#7246:}
Figure~\ref{fig_issue6} shows the result output by \name for Issue \#7246.
The reason for this slowdown is that Wasmtime incorrectly sinks a subnormal floating-point multiplication, which is originally located outside the loop, into the loop during compilation.
As shown in Figure~\ref{fig_issue6}, the multiplication in line 9 appears before the loop that begins at line 11.
However, the compiled x86-64 code places this multiplication inside the loop (line 28).
This is relatively easy to observe, and indeed, the developers initially identified it as the cause of the slowdown.

However, \name's result revealed that this alone is not sufficient to explain the issue.
The mutant generated by \name produces identical machine code to the original, including the same incorrectly hoisted multiplication inside the loop.
Yet, the original program runs 28x slower than the mutant.
The only difference lies in the operand value of the multiplication, which is stored in memory at address \textit{\%rip+0x2c} (line 28): in the original program, the value is \textit{1}, while in the mutant it is \textit{0}.
The other operand is loaded into \textit{\%xmm3} from address 0 in linear memory (lines 2–6 and 20–23) and holds a subnormal floating-point number, \textit{4.94066e-324}.
Thus, the original program performs \textit{1.0 * 4.94066e-324}, while the mutant performs \textit{0.0 * 4.94066e-324}.
The value \textit{4.94066e-324} falls within the subnormal range of floating-point numbers (\ie, between \textit{4.9e-324} and \textit{2.23e-308}), which causes the multiplication to be handled via a much slower hardware fallback path~\cite{marc2015subnormal}.
Such operations may be up to 100x slower than multiplications involving normal-range operands.
In contrast, CPUs typically implement a fast path for zero multiplication, allowing \textit{0.0 * 4.94066e-324} to complete in just a few cycles (normal multiplications also take only a few cycles to execute).
Therefore, the performance degradation is not merely due to the extra instructions in the loop, but specifically due to the sinking of a \textit{subnormal multiplication} that is computationally expensive.

Initially, the Wasmtime developers assumed that the slowdown was caused by the redundant instructions introduced by the incorrect code sinking.
However, \name demonstrated that modern CPUs can efficiently handle extra instructions in most cases.
Instead, certain rare instructions, like subnormal floating-point operations, can cause significant slowdowns and must be carefully considered during runtime optimization or user system development.
After we shared our findings, one developer commented:
``Fascinating! Not what I would have expected but helps explain the results for sure!''

\subsection{Contributions of Individual Components}\label{eval_rq3}

To demonstrate the effectiveness of the oracle runtime in selecting suitable mutated programs, we conducted a comparison in which all mutants were ranked using only the \texttt{perfDiffScore} from Algorithm~\ref{algo_scoring}, \ie, without incorporating the \texttt{funcSimScore} computed from execution time on the oracle runtime.
Under this setting, the mutants that successfully isolated the \codes for Issues \#7085, \#8573, \#8571, \#8706, \#9590, \#7246 were ranked 9, 36, 25, 98, 24, and 18, respectively.
Manually checking and filtering out all higher-ranked mutants would be highly inefficient.
Moreover, the top-10 mutants ranked solely by \texttt{perfDiffScore} had near-zero \texttt{funcSimScores}, indicating that their functionalities are significantly different from the original programs.
These results demonstrate that the oracle runtime plays a crucial role in filtering out invalid candidates and improving the efficiency when selecting suitable mutants for isolation.

To evaluate the effectiveness of \name’s fine-grained mutation strategy, we compared it against \textit{wasm-mutate}\cite{wasmmutate}.
Mutation methods for \wasm~\cite{Wenxuan2024Wapplique,shangtong2024WASMaker,jiang2025distinguishability,zhou2025lwdiff} typically extract code snippets from real-world \wasm programs and insert or substitute them into seed programs to generate mutants.
These methods often introduce substantial functional differences from the original program, making them unsuitable for isolating \codes for performance issues.
Therefore, we did not include them as baselines.
\textit{wasm-mutate} supports semantically equivalent transformations (via the \textit{--preserve-semantics} flag), which can generate mutated programs functionally similar to the original one as much as possible.
Thus, we selected it as a baseline for comparison.
Specifically, for each of Issues \#7085, \#8573, \#8571, \#8706, \#9590, \#7246, we used \textit{wasm-mutate} to generate 500 mutated programs.
After deduplication, around 350 unique programs remained per issue.
We then ranked these programs using Algorithm~\ref{algo_scoring}.

Overall, we found that the \codes isolated from the top-10 mutants generated by \textit{wasm-mutate} failed to explain the root causes in all six cases.
For Issues \#8573, \#8571, and \#9590, only 4, 2, and 2 valid mutants (\ie, those with both high \texttt{funcSimScore} and \texttt{perfDiffScore}) were found, respectively.
We further examined these valid mutants and the top-10 candidates for Issues \#7085, \#8706, and \#7246.
We observed that \textit{wasm-mutate} often modifies programs at coarse granularity: it removes and inserts multiple \wasm instructions (typically around a dozen), and sometimes alters control-flow structures or introduces new blocks.
These extensive changes result in substantial differences in the generated machine code, making it difficult to attribute performance anomalies to specific causes.
For example, the average number of modified \wasm instructions was 8.6, 19.5, 33.0, 19.2, 24.5, and 5.3 for Issues \#7085, \#8573, \#8571, \#8706, \#9590, \#7246, respectively; the corresponding average machine instruction differences were 5.0, 11.0, 12.5, 32.3, 14.0, and 1.7.
In contrast, \name applies mutations at the granularity of individual \wasm instructions.
It explicitly excludes control-flow instructions and minimizes the insertion of new instructions.
Across Issues \#7085, \#8573, \#8571, \#8706, \#9590, \#7246, \name modified only 1–2 \wasm instructions per mutant, enabling precise isolation of the \codes in each case.

\section{Discussion and Limitations}\label{sec_dis}

In our experiments, we evaluate \name on a limited number of performance issues.
As a tool designed to assist performance debugging, \name aims to address diverse, real-world, and previously unknown issues that have not been diagnosed before.
Following common practice in prior debugging work~\cite{xiang2023relational,yongle2019the}, we adopted a case-based evaluation methodology, \ie, we applied the tool to a limited number of real-life issues (since such issues with unknown root causes are, by nature, rare).
Our evaluation demonstrates that \name meets its design goal: it successfully helped diagnose 10 out of 12 performance issues, including 6 previously unknown cases.

Besides, while \name can identify suboptimal instructions responsible for abnormal performance, further analysis is still required by developers to understand these instructions and infer the final root causes.
Nevertheless, identifying the \codes is often the first and daunting step in performance debugging.
Our evaluation shows that \name effectively achieves this step and provides significant insights that help developers analyze the underlying causes.

Last, \name takes less than 3 hours to isolate the \codes for each performance issue.
However, reducing the original bug-inducing \wasm program using \textit{wasm-reduce}~\cite{wasmreduce} takes an average of 11 hours per case, despite \textit{wasm-reduce} being the most commonly used reduction tool for \wasm.
Improving the efficiency of program reduction for \wasm remains an important direction for our future work.





\section{Related Work}\label{sec_re}

\textbf{\wasm Runtimes Performance.}
High performance is a critical design goal for \wasm, especially as it has been proposed as a secure and lightweight sandboxing mechanism in various application scenarios, \eg, cloud computing~\cite{simon2020Faasm,philipp2022pushing,Phani2020Sledge}.
Many studies have evaluated the performance of \wasm by comparing it with native code~\cite{abhinav2019not} and JavaScript~\cite{wang2021empowering} across different \wasm runtimes, including both browser-based~\cite{yan2021understanding} and standalone environments~\cite{spies2021evaluation}.
Recently, Jiang~\etal~\cite{jiang2023revealing} demonstrated that runtime bugs in \wasm runtimes can cause significant performance degradation when hosting cloud services.
Several works~\cite{jiang2023revealing,jiang2025distinguishability,Wenxuan2024Wapplique} focused on performance testing of \wasm runtimes.
While these existing efforts have successfully detected performance issues, none have addressed the problem of diagnosing the root causes of such issues within \wasm runtimes.
To the best of our knowledge, this work is the first to study performance issue diagnosis in \wasm runtimes.

\noindent \textbf{Performance Debugging.}
Performance debugging is widely known as a challenging and elusive task, and many studies have investigated how to facilitate this process.
Profiling tools~\cite{url-perf,url-vtune} help identify \textit{where} performance anomalies occur by reporting functions that take the most execution time.
Causal analysis~\cite{aguilera2003performance,weng2021argus,ravindranath2012appinsight,zhao2016non,barham2004using} helps to understand performance issues by identifying chains of runtime events across multiple functions during abnormal executions, starting from the point where the abnormal symptom is observed (\eg, a hot function reported by the profiler).
Other studies~\cite{xiang2023relational,linhai2014statistical,song2017performance,su2019redundant} aim to directly identify the root causes of performance issues within the buggy software.
However, these state-of-the-art approaches are primarily designed for performance issues arising from inefficient software implementations that lead to redundant computation.
They are not well-suited to diagnose the performance issues in \wasm runtimes, which are often caused by suboptimal compilation.
So, our work attempts to fill this gap by designing a mutation-based approach to isolate suboptimal instruction sequences, thereby facilitating further root cause analysis.

\noindent \textbf{Program Reduction.}
Program reduction aims to simplify a bug-inducing input to a minimal version that still reproduces the failure.
It is commonly applied as a preprocessing step for software debugging.
Since program reduction is an NP-hard problem with no optimal solution, existing methods focus on minimizing the size of the resulting input~\cite{andreas2002dd,ghassan2006hdd,cheng2018perses,meng2023ppr,zhou2025wdd}.
Although the machine code generated from the reduced \wasm program contains only dozens or hundreds of instructions, it is still difficult to identify the suboptimal instruction sequences within it.

\noindent \textbf{Mutation-based Program Generation for \wasm.}
Mutation-based program generation aims to produce a series of new \wasm programs by modifying given seed programs for runtime testing~\cite{Wenxuan2024Wapplique,shangtong2024WASMaker,zhou2023wadiff,zhou2025lwdiff,jiang2025distinguishability}.
To achieve high test coverage, existing approaches typically apply mutations at coarse granularity, \eg, block levels~\cite{Wenxuan2024Wapplique,jiang2025distinguishability} or function levels~\cite{zhou2025lwdiff,shangtong2024WASMaker}.
Some other techniques mutated \wasm programs at the byte level without considering semantics~\cite{AFLplusplus-Woot20,zhou2023wadiff}, often resulting in a large number of invalid mutants.
Overall, these approaches fail to produce valid mutants that are functionally similar to the original programs, which is necessary in our scenarios.
Therefore, our work proposes a fine-grained and type-aware \wasm mutation strategy to generate suitable mutants for isolating the \codes responsible for performance issues.

\section{Conclusion}\label{sec_con}

This paper presents the first study on diagnosing performance issues in \wasm runtimes.
We propose a mutation-based approach, \name, to accurately identify the suboptimal instruction sequences responsible for abnormal performance, thereby facilitating root cause analysis.
Our evaluation on 12 real-world performance issues demonstrates the effectiveness of \name in localizing the exact causes of abnormal performance.
In particular, \name successfully helped infer the root causes for 6 previously unknown issues.

\begin{acks}
This work is supported by the National Natural Science Foundation of China (Project No. 62572127).
\end{acks}

\bibliographystyle{ACM-Reference-Format}
\bibliography{reference}

\end{document}